\documentclass[namedreferences]{SolarPhysics}
\usepackage[optionalrh]{spr-sola-addons}
\usepackage{graphicx}

\begin{document}

\begin{opening}
\title{THE HEMISPHERIC ASYMMETRY OF SOLAR ACTIVITY DURING THE TWENTIETH
CENTURY AND THE SOLAR DYNAMO}

\author{ASHISH \surname{GOEL}\email{ashishgoel@iitb.ac.in}}
\institute{Department of Physics, Indian Institute of Technology, Bombay - 400076}
\author{ARNAB RAI \surname{CHOUDHURI}\email{arnab@physics.iisc.ernet.in}} 
\institute{Department of Physics, Indian Institute of Science, Bangalore-560012}
\runningtitle{Hemispheric asymmetry and solar dynamo}
\runningauthor{A.\ GOEL AND A.\ R.\ CHOUDHURI}
\begin{abstract}

We believe the Babcock--Leighton process of poloidal field
generation to be the main source of irregularity in the solar cycle.
The random nature of this process may make the poloidal field in one
hemisphere stronger than that in the other hemisphere at the end of
a cycle. We expect this to induce an asymmetry in the next sunspot
cycle. We look for evidence of this in the observational data and
then model it theoretically with our dynamo code. Since actual polar
field measurements exist only from 1970s, we use the polar faculae
number data recorded by Sheeley (1991) as a proxy of the polar field
and estimate the hemispheric asymmetry of the polar field in different
solar minima during the major part of the twentieth century. This
asymmetry is found to have a reasonable correlation with the asymmetry
of the next cycle. We then run our dynamo code by feeding information
about this asymmetry at the successive minima and compare with observational
data. We find that the theoretically computed asymmetries of different
cycles compare favourably with the observational data, the correlation
coefficient being 0.73. Due to the coupling between the two hemispheres,
any hemispheric asymmetry tends to get attenuated with time. The hemispheric
asymmetry of a cycle either from observational data or from theoretical
calculation statistically tends to be less than the asymmetry in the
polar field (as inferred from the faculae data) in the preceding minimum.
This reduction factor turns out to be 0.38 and 0.60 respectively in
observational data and theoretical simulation.
\end{abstract}
\end{opening}

\def\Rs{R_{\odot}}

\section{Introduction}

Although solar activity appears reasonably symmetric in the two hemispheres
after short-term variations are averaged, some cycles have been known to
be stronger in one hemisphere.  The aim of the present paper is to analyze
the asymmetries of solar cycles during the twentieth century and then
to simulate these asymmetries with a solar dynamo model.

The solar magnetic cycle is believed to be produced by a flux transport
dynamo operating in the sun's convection zone (Wang, Sheeley, and Nash,
1991; Choudhuri, Sch\"ussler,
and Dikpati, 1995;
Durney, 1995; Dikpati and Charbonneau, 1999; Nandy and Choudhuri, 2001,
2002; K\"uker, R\"udiger, and Schultz, 2001; Guerrero and Mu\~noz, 2004). 
Fairly sophisticated
models of the solar dynamo to explain various regular features of
the solar cycle have been constructed. There is, however, not yet
a convergence on the values of important parameters. 
In the model of Chatterjee, Nandy, and Choudhuri (2004), the value of turbulent
diffusivity for the poloidal field in the interior of the solar convection
zone is taken to be $2.4\times10^{12}$ cm${^{2}}$ s$^{-1}$. On
the other hand, Dikpati and Gilman (2006) take a value about 50 times
smaller.

In order to model the hemispheric asymmetry,
we need to understand how the irregularities
of the solar cycle arise in the flux transport dynamo theory.
We believe that the stochastic fluctuations in the dynamo
process give rise to the irregularities (Choudhuri, 1992).
Choudhuri, Chatterjee, and Jiang (2007) identify the
Babcock--Leighton process of the production of poloidal field as the
main source of randomness in the solar dynamo, whereas other aspects
of the dynamo process are assumed to be deterministic. In the Babcock--Leighton
process, the poloidal field is produced from the decay of tilted bipolar
sunspots. The tilt of bipolar sunspots is caused by the Coriolis force
acting on the rising flux tubes (D'Silva and Choudhuri, 1993), whereas
buffeting of the flux tubes by convective turbulence causes a scatter
in the tilt angles around the average given by Joy's law (Longcope
and Choudhuri, 2002). Because of this scatter in tilt angles, the Babcock--Leighton
process appears not to be a deterministic process. Observational data,
as plotted in Figure~3 of Jiang, Chatterjee and Choudhuri (2007), 
also indicate that
the polar field produced at the end of a cycle is not correlated with
the strength of the cycle. On the other hand, Dikpati and Gilman (2006)
use the sunspot area data as the source function for the poloidal
field, which amounts to assuming the Babcock--Leighton process to
be fully deterministic and which is incorrect in our opinion. 
Dikpati and Gilman (2006) have predicted that the next cycle~24
will be 30--50\% stronger than the last cycle, which is at variance
with the prediction of Choudhuri, Chatterjee, and Jiang (2007)
and Jiang, Chatterjee, and Choudhuri (2007) that it will be
30--35\% weaker.

Although the polar field produced at the end of a cycle is not correlated
with the strength of the cycle, observational data show that the strength
of the cycle is correlated quite well with the polar field at the
preceding minimum. This is seen in Figure~2 of Jiang, Chatterjee, and Choudhuri (2007).
In fact, Schatten et al.\ (1978) proposed long ago that the strength
of the polar field at a solar minimum can be used to predict the strength
of the next cycle. Svalgaard, Cliver, and Kamide (2005) and Schatten (2005) have
used the weakness of the present polar field to predict that the next
cycle~24 will be weak. Jiang, Chatterjee, and Choudhuri 
(2007) showed that only a reasonably
high value of turbulent diffusivity can give rise to the observed
correlation between the polar field at the minimum and the strength
of the next cycle. How this correlation arises is explained through
Figure~1 of Jiang, Chatterjee, and Choudhuri (2007). If the diffusivity is high, then
the poloidal field generated at the solar surface by the Babcock--Leighton
process diffuses to the tachocline in a few years. Since the next
cycle is caused by the toroidal field produced 
from this poloidal field in the tachocline by
differential rotation, it is obvious that the next cycle would appear
correlated with the preceding polar field which is formed by the poleward
advection of the poloidal field due to meridional circulation. On
the other hand, if the diffusivity is low, then the poloidal field
produced at the surface cannot diffuse to the tachocline and has to
be carried to the tachocline by the meridional circulation. This takes
about 20 years so that a particular cycle is not correlated with the
polar field in the immediately preceding minimum. Dikpati and Gilman
(2007) could predict a strong cycle after a minimum with a weak polar
field only because they used a low diffusivity. This would never be
possible in a high-diffusivity model. Jiang, Chatterjee, and Choudhuri (2007; \S5)
provided several independent arguments why the diffusivity is likely
to have the higher value which they assumed. Yeates, Nandy, and Mackay (2007)
have recently carried out a thorough study of the effects of diffusivity
on a fluctuating dynamo and have confirmed the findings of Jiang,
Chatterjee, and Choudhuri (2007).

If the Babcock--Leighton process of poloidal field generation is the
source of randomness in the solar dynamo, then a theoretical model
based on mean field equations has to be corrected by feeding the actual
value of the observed polar field at the solar minimum (Choudhuri,
Chatterjee, and Jiang, 2007). 
Since reliable polar field measurements are available only from mid-1970s,
Choudhuri, Chatterjee, and Jiang (2007) and Jiang, Chatterjee, and
Choudhuri (2007) attempted to model
only the last three solar cycles. As these last three cycles were only weakly
asymmetrical between the hemispheres, they are not particularly
convenient in studying the physics of hemispheric asymmetry, although
Jiang, Chatterjee, and Choudhuri (2007) presented some calculations
of hemispheric asymmetry.
Jiang, Chatterjee, and Choudhuri (2007) pointed
out two other works which provide proxies for the polar field at earlier
minima: (i) the polar faculae numbers analyzed by Sheeley (1991);
and (ii) large-scale magnetic moments obtained by Makarov et al.\ (2001)
from the positions of dark filaments. While Jiang, Chatterjee, and
Choudhuri (2007)
carried out some correlation analyses based on these data, they were
not used in dynamo modelling. Since Sheeley (1991) has provided
both the north and south polar faculae numbers during 1906--1990,
we can use this to estimate the asymmetries in the polar field during
the various solar minima of the twentieth century. Jiang, Chatterjee,
and Choudhuri (2007) stressed the fact that polar fields inferred from
the faculae data may not always be reliable. Since it is still the best
that we can do to model the asymmetries of earlier cycles, 
it is instructive to see
what we get from this approach.

The randomness of the Babcock--Leighton process may give rise
to a stronger poloidal field in one hemisphere compared to the
other.  Just as the polar field at the minimum gives an
indication of the strength of the next cycle, we may expect that a
hemispheric asymmetry in the polar field may be indicative of a hemispheric
asymmetry in the solar activity during the next cycle. We find a reasonably
good correlation in the observational data. The theoretical dynamo
model with our assumed value of diffusivity reproduces this correlation
qualitatively. In spite of a large scatter in the data,  
we can clearly see some interesting patterns. 

We present a discussion of hemispheric asymmetry seen in the observational
data in \S2. Then \S3 presents calculations from our dynamo model
in which magnetic field values in the two poles during the minima
are fed. The theoretical results of hemispheric asymmetry are discussed
in \S4. Our conclusions are summarized in \S5.

\section{Observational data}

\begin{table}
\begin{tabular}{|l|r|r|r|r|r|r|}
\hline 
Cycle &
\multicolumn{3}{c|}{Polar faculae number}&
\multicolumn{3}{c|}{Total sunspot area}\\
Number &
\multicolumn{3}{c|}{at beginning of cycle}&
\multicolumn{3}{c|}{during the cycle}\\
\cline{2-7}
&
$F_{N}$ &
$F_{S}$ &
$F_{{AS}}$ &
$A_{N}$ &
$A_{S}$ &
$A_{{AS}}$ \\
\hline 
15&
28.26 &
31.59 &
-0.0566 &
43331.9 &
35689.1 &
0.096719 \\
16&
53.85&
49.43&
0.0428&
46509.0&
39079.8&
0.086801 \\
17&
25.19&
30.62&
-0.0973&
60023.8&
59649.7&
0.003126 \\
18&
51.51&
33.03&
0.2186&
74255.4&
70292.3&
0.027417 \\
19&
64.76&
44.13&
0.1895&
105511.0&
73887.7&
0.176274 \\
20&
66.19&
36.89&
0.2842&
69387.4&
49101.1&
0.171209 \\
21&
24.54&
29.18&
-0.0864&
75077.2&
77623.3&
-0.016674 \\
22&
23.62&
26.28&
-0.0533&
63790.6&
72407.2&
-0.063265 \\
\hline
\end{tabular}

\caption{Polar faculae numbers and total sunspot areas in two
hemispheres during the various cycles.}
\end{table}

We use Figure~1 of Sheeley (1991) to estimate the north polar faculae
number ($F_{N}$) and the south polar faculae number ($F_{S}$) at
successive solar minima. The values of $F_{N}$ and $F_{S}$ at the
beginnings of various cycles are listed in Table~1 along with the
asymmetry factor \begin{equation}
F_{AS}=\frac{F_{N}-F_{S}}{F_{N}+F_{S}}\end{equation}
 It should be noted that the polar faculae number plotted in Figure~1
of Sheeley (1991) is often noisy near the solar minima when this number
has maximum values. So, when using $F_{AS}$ as a proxy for the asymmetry
in the polar field, the possibility of significant errors should be
kept in mind. Since actual measurements of polar field from WSO were
available since 1976, Sheeley (1991) presented a comparison of actual
polar field values and the faculae numbers during the period when
both types of data were available (see Figures~2 and 3 in his paper).
While the correlation between the two appears reasonably good, it
is certainly not extremely tight. Jiang, Chatterjee, and Choudhuri (2007) pointed out
that the polar field inferred the faculae number data of Sheeley (1991)
did not always agree with the polar field inferred from the parameter
$A(t)$ computed by Makarov et al.\ (2001) from the positions of
dark filaments.

%\begin{figure}
%\centering{\includegraphics[width=6cm]{fig1.eps}}
%\caption{}
%\end{figure}

\begin{figure}
\centering{\includegraphics[width=12cm]{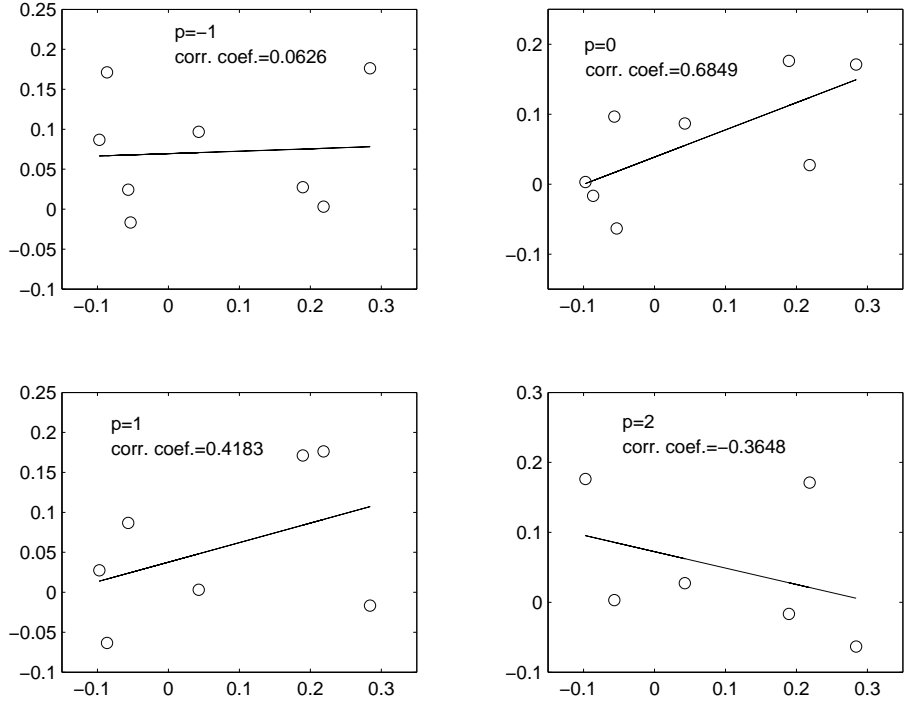}}

\caption{The observed asymmetry in sunspot area $A_{AS}$ of cycle $n+p$ is
plotted against the polar faculae asymmetry $F_{AS}$ at the beginning
of the cycle $n$.}
\end{figure}

To compute asymmetries of sunspot cycles, we use the sunspot area
data from the archive of Royal Greenwich Observatory available at
the website:

http://solarscience.msfc.nasa.gov/greenwch.shtml

\noindent Monthly averages of daily sunspot areas for the northern
and southern hemispheres are available at this website. We add up
the monthly sunspot areas over one sunspot cycle to get a `total'
sunspot area during the cycle in one hemisphere. Let us denote these
`total' sunspot areas in the two hemispheres summed over sunspot cycles
by $A_{N}$ and $A_{S}$. Table~1 also lists the values of $A_{N}$
and $A_{S}$ for various sunspot cycles along with the asymmetry factor
\begin{equation}
A_{AS}=\frac{A_{N}-A_{S}}{A_{N}+A_{S}}\end{equation}
Figure~1 now plots the sunspot area asymmetry $A_{AS}$ of cycle~$n+p$
against the polar faculae asymmetry $F_{AS}$ at the beginning of
the cycle~$n$. Plots are shown for four values of $p$: $-1$, 0, 1
and 2. The lack of correlation in the plot for $p=-1$ suggests that the
asymmetry of the cycle does not determine the asymmetry of the polar
faculae at the end of the cycle. We have the best correlation when
$p=0$. The correlation becomes somewhat weaker for $p=1$ and virtually disappears
for $p=2$. The message is quite clear. The asymmetry of the poloidal
field produced at the end of a sunspot cycle is the major factor determining
the asymmetry of the next cycle. This would be possible only if the
information about the poloidal field asymmetry at the solar surface
can be communicated within a few years ($\approx$ 5 years) to the
tachocline which is the breeding ground for the sunspots in the next
cycle. As argued by Jiang, Chatterjee, and Choudhuri (2007) and confirmed by Yeates,
Nandy, and Mackay (2007), this requires a diffusivity of the order $2.4\times10^{12}$
cm${^{2}}$ s$^{-1}$ as used by Chatterjee, Nandy, and Choudhuri 
(2004) and Choudhuri, Chatterjee, and Jiang
(2007). If the diffusivity is assumed to be 50 times smaller
as in Dikpati and Gilman (2006), then
diffusion cannot carry an information from the solar surface to the
tachocline in a reasonable time. This has to be done by the meridional
circulation, which has an advection time of about 20 years. On using
such a low value of diffusivity in their numerical simulations, Charbonneau
and Dikpati (2000) found that the polar field at the beginning of
a cycle $n$ had the maximum correlation with the strength of the
cycle $n+2$, there being virtually no correlation with the cycle
$n$ (see their Figure~9).

The `memory' of the solar cycle can be estimated from Figure~1. Given
the fact that the correlation becomes weaker from the $p=0$ to the
$p=1$ case and disappears in the $p=2$ case, the `memory' is expected
to be of order 15--20 years. This is completely consistent with Figure~2
of Choudhuri, Chatterjee, and Jiang (2007), where we see that the effect of a sudden
disturbance persists for about 15--20 years. While it may be unlikely
that all the parameters used by Chatterjee, Nandy, and Choudhuri (2004), 
Choudhuri, Chatterjee, and Jiang (2007)
and Jiang, Chatterjee, and Choudhuri (2007) have the exactly correct
values, the values of quantities like diffusivity probably have been
chosen correct within a factor of 2 or 3, since `memory' from this
model is in good agreement with the limited observational data that
we have. If the `memory' is longer than a cycle, then the randomness
introduced by the Babcock--Leighton process at the end of a cycle
does not erase all the effects of the previous cycle completely. Bushby
and Tobias (2007) have argued against very long memories in a complex
nonlinear system like the solar dynamo. On the other hand, Charbonneau,
Beaubien, and St-Jean (2007) suggested that the `even-odd' effect in the solar
cycle is caused by period doubling, which would imply a memory which
is at least as long as what we are suggesting.

The last important point to note in the observational data is that
the correlation line for the $p=0$ case in Figure~1 has a slope
of 0.38. Even if the polar field asymmetry at a minimum is the primary
cause of the asymmetry in the next cycle, it seems that the asymmetry
in the cycle is statistically expected to be only 0.38 times the polar
asymmetry. In other words, the asymmetry tends to get reduced as the
cycle progresses. Chatterjee and Choudhuri (2006) studied the coupling
between the two hemispheres and showed that, for a dynamo with high
diffusivity, the two hemispheres remain coupled even after the introduction
of asymmetries. So we expect that the hemispheric asymmetries continously
get washed away until the randomness in the Babcock--Leighton process
towards the end of a cycle
creates fresh asymmetries.

One may wonder whether we would get plots similar to what we see in
Figure~1 when we try to correlate the total polar faculae number
$F_{N}+F_{S}$ at the beginning of cycle $n$ with the `total' sunspot
area $A_{N}+A_{S}$ during cycle $n+p$. Some plots of this kind are
shown in Figs.~2 and 3 of Jiang, Chatterjee, and Choudhuri (2007). When we estimate
the polar field at the beginning of cycle $n$ from the value of $A(t)$
computed by Makarov et el.\ (2001) and correlate it with the `total'
sunspot area of cycle $n+p$, we get plots very similar to the plots
in Figure~1. However, when we carry out such an exercise by taking
$F_{N}+F_{S}$ as a proxy of the polar field, we do not get very clear
plots. Even in the case $p=0$, we do not find a strong correlation.
It is intriguing that we get the interesting plots of Figure~1 by
correlating the asymmetries $F_{{AS}}$ and $A_{{AS}}$, but
we do not get such plots when we try to correlate $F_{N}+F_{S}$ and
$A_{N}+A_{S}$ for different cycles. We have no proper explanation
for this. We merely record this fact here. One possibility is that
$F_{{AS}}$ is a better proxy for the polar field asymmetry than
$F_{N}+F_{S}$ is a proxy for the average polar field. We, however,
cannot think up a good reason why this should be the case.

\section{The numerical dynamo model}

We now carry out an analysis of the asymmetry in solar activity on
the basis of the standard dynamo model presented by Nandy and Choudhuri
(2002) and Chatterjee, Nandy, and Choudhuri (2004). The basic equations for the
standard axisymmetric $\alpha\Omega$ solar dynamo model can be found
in Chatterjee, Nandy, and Choudhuri (2004). In order to solve these governing equations,
we make use of the solar dynamo code SURYA developed by the research
group at the Indian Institute of Science. This code and a detailed
guide (Choudhuri, 2005) can be availed upon request by sending an e-mail
to Arnab Rai Choudhuri (email address: arnab@physics.iisc.ernet.in).
The code SURYA has been the basis for dynamo calculations
presented in several papers
(Chatterjee, Nandy, and Choudhuri, 2004; Choudhuri, Chatterjee, and
Nandy, 2004; Chatterjee and Choudhuri, 2006; Choudhuri, Chatterjee,
and Jiang, 2007; Jiang, Choudhuri, and Wang, 2007; Jiang, Chatterjee,
and Choudhuri, 2007; Yeates, Nandy, and McKay, 2007).

As discussed earlier, the Babcock-Leighton process of poloidal field
generation from the decay of tilted bipolar sunspot pairs involves
randomness. Hence, in order to analyze the irregularities of the solar
cycles, we have to force-feed the observational data for the poloidal
field at the solar minima. To accomplish this, Choudhuri, Chatterjee, and Jiang (2007)
adopted the following method. Cycle~22 was chosen as the average cycle
and the observed value of the polar field at a solar minimum was divided
by the value of the polar field at the beginning of cycle~22 to arrive
at a numerical factor $\gamma$. This constant $\gamma$ is essentially
a measure of the observed poloidal field at a solar minimum. Now let
$\overline{A_{\rm min}}$ be the amplitude of the scalar function $A(r,\theta)$
which gives the poloidal field at the minima of a relaxed solution
of the dynamo code. The code was stopped at successive minima,
when $A(r,\theta)$ above $r>0.8R\odot$ would be multiplied by a
constant factor such that its amplitude becomes equal to $\gamma\overline{A_{\rm min}}$,
where $\gamma$ is the numerical factor appropriate for that minimum.
Values of $A(r,\theta)$ below $r<0.8R\odot$ were left unchanged
to ensure that only the poloidal field created in the previous cycle
would be updated, but any poloidal field created in still earlier
cycles which may be present at the bottom of the convection zone was
not changed. Choudhuri, Chatterjee, and Jiang (2007) used a single $\gamma$ for
the whole Sun at every minimum. On the other 
hand, Jiang, Chatterjee, and Choudhuri (2007)
used a function $\gamma(\theta)$ of the latitude obtained from WSO
data of poloidal field at different latitudes. We now follow the procedure
of assigning two different $\gamma_{N}$ and $\gamma_{S}$ for the
two hemispheres obtained from the north and south polar faculae numbers
during the minima. If we again take the cycle~22 as an average cycle,
we see in Table~1 that the average value of polar faculae number
(i.e.\ the average of north and south poles) at the beginning of
that cycle was 24.95. Dividing the numbers in the second and third
columns of Table~1 by this, we get the values of $\gamma_{N}$ and
$\gamma_{S}$.

On the basis of this methodology, we carry out simulations for cycles
15--22 by updating the poloidal field at the minima with the help
of the polar faculae number data of Sheeley (1991). Before presenting
the results of asymmetry, we show a theoretical sunspot number plot
in Figure~2 along with the observational data. As already pointed
out by Choudhuri, Chatterjee, and Jiang (2007) and 
Jiang, Chatterjee, and Choudhuri (2007), the absolute
value of the theoretical sunspot number does not have a particular
physical significance. So we have scaled it appropriately to produce
a good fit with the observational data. We found that the
theoretically calculated cycles vary
in duration slightly if we feed the poloidal field data at the minima
by our procedure. It is believed that the duration of a cycle is set
by the time scale of the meridional circulation (Charbonneau and Dikpati,
2000; Hathaway et al., 2003), and helioseismology gives us information
about the variation of meridional circulation only from 1996 onwards.
Most probably, it is the variation of meridional circulation with
time which is the primary cause of variation in the observed durations
of cycles. Since we do not have any information of meridional circulation
variation at earlier times, we take the meridional circulation to
be constant in our model and do not try to match the observed variation
of cycle durations. The total duration of cycles~15--22 in our theoretical
model turned out to slightly longer than the observed duration. We
had to shrink the time axis in the theoretical model by a factor 0.86
to produce Figure~2.

\begin{figure}
\centering{\includegraphics[width=12cm,height=6cm]{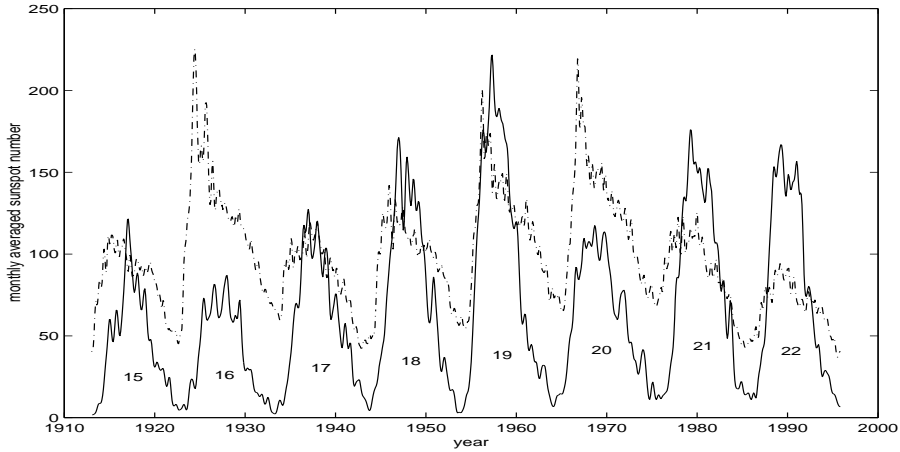}}

\caption{The solid line represents the
monthly averaged sunspot numbers from observation, while the dash-dotted
line represents the theoretical monthly averaged sunspot number calculated
by feeding the polar faculae data of Sheeley (1991) in the dynamo code.}
\end{figure}

It was mentioned by Chatterjee, Nandy, and Choudhuri (2004) that one of the limitations
of their model (which we use here) is that the theoretical sunspot
number at the minima remained significantly non-zero. We see in Figure~2
that there is no good match between theory and observations during
the solar minima. This was the case in the results of Choudhuri, Chatterjee, and
Jiang (2007) and Jiang, Chatterjee, and Choudhuri (2007) as well. The fits between theory
and observations during the maxima of most of the cycles seem reasonable,
except the two weak cycles~16 and 20, as well as the last
cycle~22. The two weak cycles~16 and 20 correspond
to the two data points in Figure~2(b) of Jiang, Chatterjee, and Choudhuri (2007) which
are quite a bit away from the correlation line. As pointed out by
Jiang, Chatterjee, and Choudhuri (2007), these two weak cycles were preceded by fairly
high values of polar faculae number suggesting a strong polar field
in the previous minimum, whereas the polar field inferred from the
value of $A(t)$ as computed by Makarov et al.\ (2007) is on
the lower side.

For the sake of comparison, we also carried out a calculation of cycles
16--23 by feeding the polar field data at the minima inferred from
the values of $A(t)$ given by Makarov et al.\ (2007). The result
is shown in Figure~3. Note that, for this calculation, a single value
of $\gamma$ was used at each minimum, which was taken to be proportional
to $A(t)$ at that minimum. 
We see that the fit between theory and observation is better in this
case. This was expected because the correlation plot given in Figure~2(a)
of Jiang, Chatterjee, and Choudhuri (2007) based on the data of Makarov et al. (2001)
shows a tighter correlation than the correlation plot given in Figure~2(b)
based on the polar faculae data of Sheeley (1991).

\begin{figure}
\includegraphics[width=12cm,height=6cm]{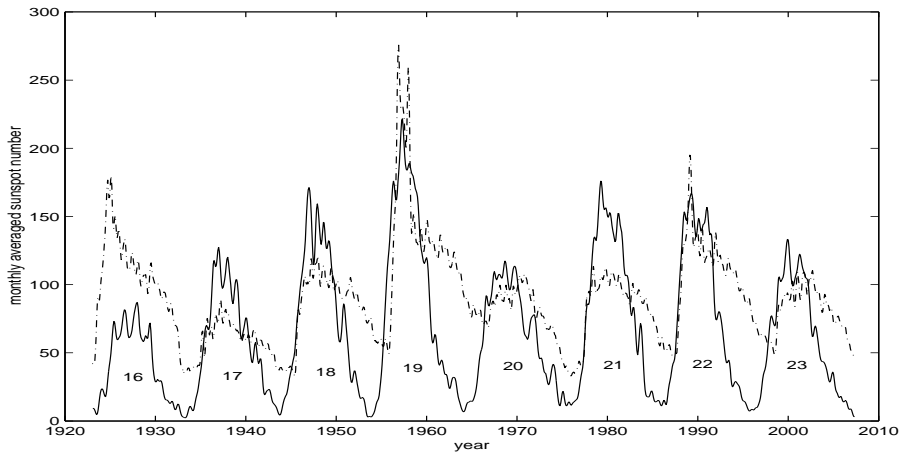}

\caption{The solid line represents
the monthly averaged sunspot numbers from observation, while the dash-dotted
line represents the theoretical monthly averaged sunspot number calculated
by feeding the polar field value inferred from the data of Makarov et al.\
(2001) in the dynamo code.}
\end{figure}

%\begin{figure}
%\includegraphics[width=12cm]{Fig3}

%\caption{Results for data from Makarov et al. (2001). The solid line represents
%the monthly averaged sunspot numbers from observation while the dash-dot
%line represents the theoretical monthly averaged sunspot number}
%\end{figure}

%\begin{figure}
%\includegraphics[width=12cm]{Fig4}

%\caption{Results for data from Sheeley (1991). The solid line represents the
%monthly averaged sunspot numbers from observation while the dash-dot
%line represents the theoretical monthly averaged sunspot number}
%\end{figure}

\section{The asymmetry calculation}

The upper panel of 
Figure 4 shows the theoretical sunspot numbers in the two hemispheres 
from our dynamo simulation as functions
of time for the cycles~15--22. 
The theoretical curve shown in Figure~2 is nothing but the sum of
the two curves shown in Figure~4. For the sake of comparison,
the observational data of monthly 
sunspot areas in the two hemispheres as functions
of time are shown in the bottom panel of Figure~4. Both in
the theoretical and observational plots, the northern hemisphere
is found considerably more active than the southern hemisphere
during cycles~19 and 20.  These were the cycles with the strongest
asymmetry during the twentieth century.
The areas below the curves in the top panel of Figure~4
for a particular cycle give the theoretical total sunspot numbers
$N_{N}$ and $N_{S}$ in the two hemispheres for that cycle. We can
then calculate the theoretical asymmetry of a cycle in the usual way:
\begin{equation}
N_{AS}=\frac{N_{N}-N_{S}}{N_{N}+N_{S}}\end{equation}
The theoretically calculated values of asymmetry $N_{AS}$ for various
cycles is listed in Table~2, along with the values of observed asymmetry
$A_{AS}$ which were already listed in the last column of Table~1.
Then the third column of Table~2 gives the ratio of the theoretical
asymmetry to the observed asymmetry, whereas the last column lists
the difference between them. For the cycles which had sufficient
observed asymmetry (i.e.\ more than 10\%), we find this ratio to be of order
1. However, when the asymmetry is small (i.e.\ less than 10\%), it
does not have much statistical significance and sometimes the
theoretical and observational asymmetries even have opposite signs.
Only for the cycle~17 which had the weakest observed asymmetry of only
0.3\%, the ratio given in the third column of Table~2 is off from 1
by more than an order of magnitude.  However, we find in the last
column that the difference between theoretical and observed asymmetries
in this case is quite small. We conclude that our 
theoretical dynamo model produces the approximately correct value 
of asymmetry when it is sufficiently large.   

\begin{figure}
\centering{\includegraphics[width=12cm,height=10cm]{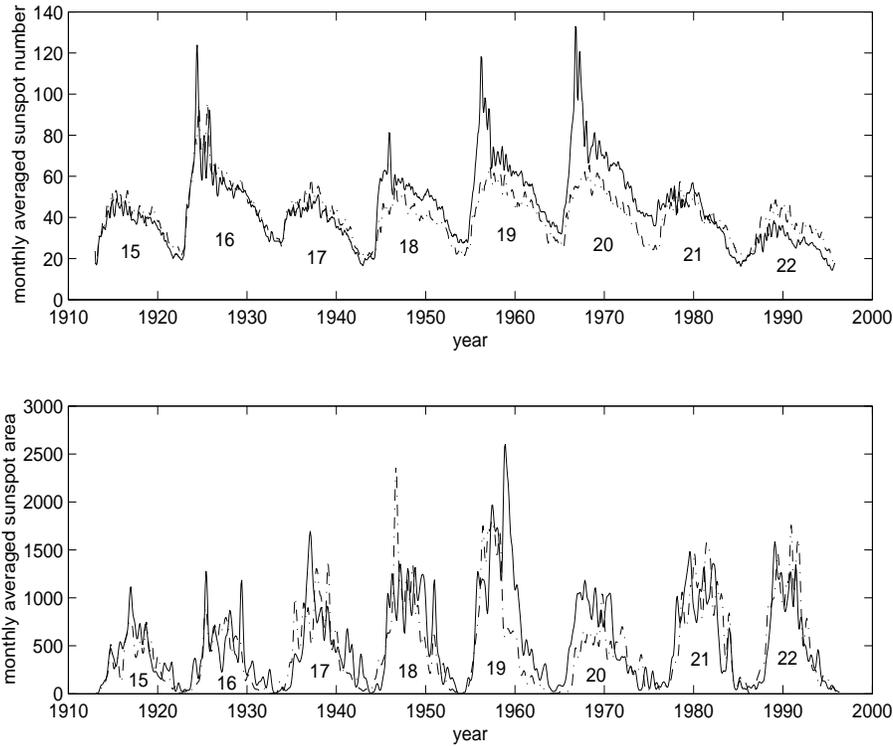}}

\caption{The top panel plots the theoretical monthly averaged sunspot 
numbers in the northern (solid line) and the southern (dash-dotted 
line) hemispheres. The bottom panel shows the observational plot for the same.}
\end{figure}

\begin{table}

\begin{tabular}{|c|c|c|c|c|}
\hline 
Cycle number&
$N_{AS}$&
$A_{AS}$&
$\frac{N_{AS}}{A_{AS}}$&
$N_{AS}-A_{AS}$\tabularnewline
\hline
15&
-0.0444&
0.096719 &
-0.4591&
-0.0411\\
16&
-0.0155&
0.086801 &
-0.1786&
-0.1023\\
17&
-0.0474&
0.003126 &
-15.1632&
-0.0505\\
18&
0.1027&
0.027417 &
3.7456&
0.0753\\
19&
0.1315&
0.176274 &
0.7460&
-0.0448\\
20&
0.1823&
0.171209 &
1.0648&
0.0111\\
21&
0.0171&
-0.016674 &
-1.0255&
0.0338\\
22&
-0.1154&
-0.063265 &
1.8241&
-0.0521\tabularnewline
\hline

\end{tabular}

\caption{Theoretical ($N_{AS}$) and observed ($A_{AS}$) asymmetries in solar activity.}

\end{table}

\begin{figure}
\centering{\includegraphics[width=10cm]{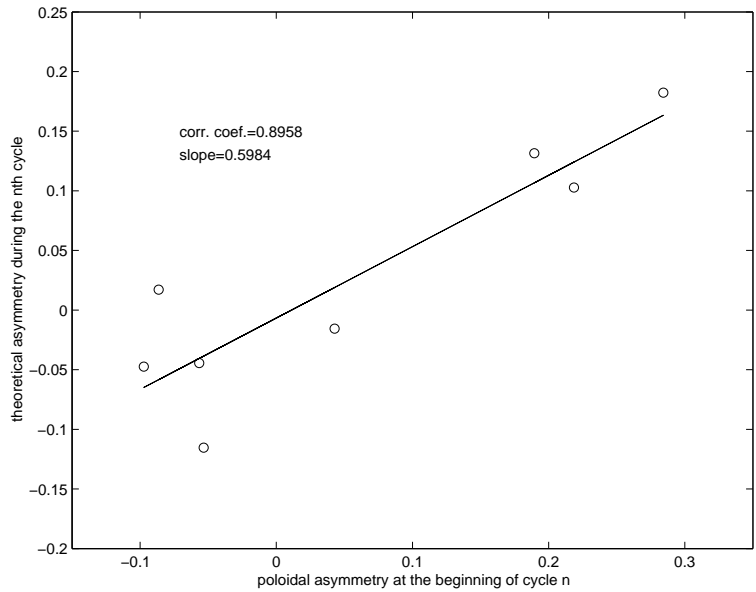}}

\caption{Theoretically calculated asymmetry during the cycle $n$ is 
plotted against the observed asymmetry in the polar faculae number at
the beginning of the cycle.}

\end{figure}

In Figure~5 we plot the theoretically calculated asymmetry $N_{AS}$
for cycle~$n$ against the asymmetry $F_{AS}$ in the polar faculae
number at the beginning of the cycle~$n$, which is essentially the
asymmetry between $\gamma_{N}$ and $\gamma_{S}$ values that have
been fed into the code.
We have to compare the theoretical Figure~5 with the corresponding
observational figure which is the plot for $p=0$ in Figure~1. Compared
to the slope 0.38 in that figure, the slope in Figure~5 has a somewhat
higher value of 0.60. We consider this to be a remarkable agreement
between theory and observations. As we pointed out in \S2, the coupling
between the hemispheres tends to reduce any asymmetry between the
hemispheres. Hence we find that the observed asymmetry $A_{AS}$ of
a cycle is less than the asymmetry $F_{AS}$ of polar faculae number
at the beginning of that cycle, which is an indication of the source
of asymmetry in the cycle. We now find that the theoretically calculated
asymmetry $N_{AS}$ of the cycle is also reduced compared to $F_{AS}$
at the beginning of the cycle and the reduction is by a factor which
is comparable to the factor we see in the observational data. We believe
that this is again an indication that parameters like diffusivity
which are responsible for the coupling between the hemispheres probably
have values in the correct ball park in our dynamo model. Figure~6
plots theoretical asymmetry $N_{AS}$ against the observational asymmetry
$A_{AS}$ for different cycles. The correlation coefficient of 0.73
is quite remarkable, judging by the fact that considerable uncertainties
are involved in using the polar faculae number as the proxy of the
polar field.

\begin{figure}

\centering{\includegraphics[width=10cm]{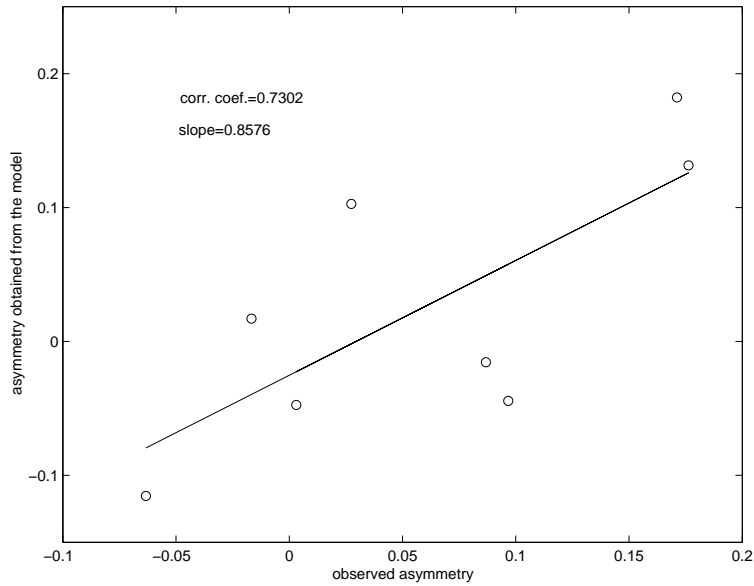}}

\caption{The theoretical asymmetry $N_{AS}$ of various cycles is plotted
against the observational asymmetry $A_{AS}$.}

\end{figure}

%

%\begin{figure}
%\includegraphics[width=6cm]{Fig5}

%\caption{Theoretical values of the asymmetry factor plotted against the observed
%asymmetry in solar activity}
%\end{figure}

%\begin{figure}
%\includegraphics[width=12cm]{Fig6}

%\caption{Theoretically calculated, monthly averaged sunspot number in the
%northern hemisphere (solid line) is superimposed over the monthly
%averaged sunspot number in the southern hemisphere (dash-dot line)}
%\end{figure}

\section{Conclusion}

During the twentieth century, some cycles had hemispheric asymmetry
larger than 17\% as seen in Table~1. It is possible that the hemispheric
asymmetry of the solar activity plays an important role in determining
the character of the solar cycle. For example, there is some observational
evidence that the there was a large hemispheric asymmetry at the time
of the onset of the Maunder minimum (Sokoloff and Nesme-Ribes, 1994)
and this asymmetry may even have played some role in inducing the
Maunder minimum (Charbonneau, 2005). However, to the best of our knowledge,
not much systematic effort has been made previously to study the asymmetry
of solar activity with the help of dynamo models.

The randomness of the Babcock--Leighton process can make the poloidal
field in one hemisphere stronger than the other and we suggest that
this induces an asymmetry in the solar cycle. We have direct poloidal
field data only from mid-1970s. Cycles from that time onwards have
been only mildly asymmetric and hence are not particularly suitable
for studying hemispheric asymmetry. Also, we need a larger data set
to draw any statistically significant conclusions. So we use the polar
faculae number reported by Sheeley (1991) as the proxy of the polar
field. In spite of uncertainties involved in this procedure, we find
that the asymmetry in the polar faculae number during a solar minimum
is correlated with the hemispheric asymmetry of the next cycle. The
correlation becomes weaker with succeeding cycles, suggesting a memory
of about 15--20 years. We point out that this type of correlation is
possible only if we assume a relatively high value of diffusivity
like $2.4\times10^{12}$ cm${^{2}}$ s$^{-1}$ (Chatterjee, Nandy,
and Choudhuri, 2004). A diffusivity of this order gives the right
kind of memory when the dynamo is subjected to a disturbance in the
poloidal field generation (Choudhuri, Chatterjee, and Jiang 2007).

When we run our dynamo code by feeding the appropriate asymmetry at
successive minima and model the sunspot cycles during the twentieth
century, we get a qualitative agreement between theory and observations.
We know that the cross-hemispheric coupling tries to reduce any asymmetry
between the two hemispheres (Chatterjee and Choudhuri, 2006). Both
in observational data and theoretical simulations, we find that the
asymmetry of a cycle statistically tends to be less than the asymmetry
in the faculae number during the preceding minimum. The reduction
factors also turn out to be comparable in the observational data and
theoretical simulation. This is quite a remarkable agreement, given
the many uncertainties involved in our analysis. Solar physicists
may have to wait for about half a century to be able to carry out
an analysis like the present analysis based on the actual measured
polar field asymmetries rather than using proxies like the polar faculae
number.  Such an analysis will be more relevant than the present
analysis, provided there will be some strongly asymmetric cycles
in the next half century.  We, however, hope that our methodology will provide
the framework for any such future analysis.

\section*{Acknowledgement}

Ashish Goel would like to thank Jawaharlal Nehru Centre for Advanced
Scientific Research for the Summer Research Fellowship Programme which
enabled him to work on this research problem.

\end{document}